\documentstyle [12pt]{article}

\newcommand{\be}{\begin{eqnarray}}
\newcommand{\ee}{\end{eqnarray}}

\begin{document}

\centerline{\bf A One Parameter Family of}
\centerline{\bf  Expanding Wave Solutions of the Einstein Equations}
\centerline{\bf That Induces an Anomalous Acceleration Into the}
\centerline{\bf Standard Model of Cosmology} 
\vspace{.3cm}

\centerline{January 5, 2009}

$$\begin{array}{ccc}
Joel\ Smoller\footnotemark[1]  &  Blake\ Temple\footnotemark[2] \nonumber \end{array}$$ 
\footnotetext[1]{Department of Mathematics, University of Michigan, Ann Arbor, MI 48109; Supported by NSF Applied Mathematics Grant Number DMS-060-3754.}

\footnotetext[2]{Department of Mathematics, University of California, Davis, Davis CA
95616; Supported by NSF Applied Mathematics Grant Number DMS-010-2493.}

\newtheorem{Theorem}{Theorem}
\newtheorem{Lemma}{Lemma}
\newtheorem{Proposition}{Proposition}
\newtheorem{Corollary}{Corollary}

\begin{abstract}

We derive a new set of equations which describe a continuous one parameter family of expanding wave solutions of the Einstein equations such that the Friedmann universe associated with the pure radiation phase of the Standard Model of Cosmology, is embedded as a single point in this family.  All of the spacetime metrics associated with this family satisfy the equation of state $p=\rho c^2/3$, correct for the pure radiation phase after inflation in the Standard Model of the Big Bang.  By expanding solutions about the center to leading order in the Hubble length, the family reduces to a one-parameter family of expanding spacetimes that represent a perturbation of the Standard Model.    We then derive a co-moving coordinate system in which the perturbed spacetimes can be compared with the Standard Model.  In this coordinate system we calculate the correction to the Hubble constant, as well as the exact leading order quadratic correction  to the redshift vs luminosity relation for an observer at the center of the expanding FRW spacetime.  The leading order correction to the redshift vs luminosity relation entails an adjustable free parameter that introduces an anomalous acceleration.   We conclude that any correction to the redshift vs luminosity relation observed after the radiation phase of the Big Bang can be accounted for, at the leading order quadratic level, by adjustment of this free parameter.  Since exact non-interacting expanding waves represent possible time-asymptotic wave patterns for conservation laws, we propose to further investigate the possibility that these corrections to the Standard Model might account for the anomalous acceleration of the galaxies, without the introduction of the cosmological constant.   
 
\end{abstract}

\section{Introduction}
\setcounter{equation}{0}

Expansion waves and shock waves are fundamental to conservation laws, because even when dissipative terms are neglected, shock-wave dissipation by itself causes non-interacting wave patterns to emerge from interactive solutions.  In this paper we construct a one parameter family of non-interacting expanding wave solutions of the Einstein equations in which the Standard Model of Cosmology, (during the pure radiation epoch), is embedded as a single point.   Our initial insight was the discovery of a new set of coordinates in which the critical Friedmann-Robertson-Walker spacetime with pure radiation sources\footnotemark[3],  \footnotetext[3]{From here on in this paper we let FRW refer to the critical ($k=0$) Friedmann-Robertson-Walker metric with equation of state $p=\rho c^2/3$, \cite{smolte}.}goes over to a standard Schwarzchild metric form, (barred coordinates), in such a way that the metric components depend only on the single self-similar variable\, $\bar{r}/\bar{t}.$  From this we set out to find the general equations for such self-similar solutions.  
In this paper we show that the PDE's for a spherically symmetric spacetime in Standard Schwarzchild coordinates (SSC) reduce, under the assumption $p=\rho c^2/3$, to a new system of three ordinary differential equations\footnotemark[4]\footnotetext[4]{As far as we are aware the only other known way the PDE's for metrics in Standard Schwarzschild Coordinates with perfect fluid sources reduce to ODE's, is the time independent case when they reduce to the Oppenheimer-Volkoff equations, \cite{wein}} in the same self-similar variable $\bar{r}/\bar{t}.$   After removing one scaling parameter and imposing regularity at the center, we prove that there exists inplicitly within the three parameter family, a continuous one parameter family of self-similar solutions of the Einstein equations that extends the FRW metric.  

Because different solutions expand at different rates,  our expanding wave equations introduce a new acceleration parameter,  and suitable adjustment of this parameter will speed up or slow down the expansion rate.  Using special properties of the metrics, we find an exact expression for the leading order (quadratic) correction to the red-shift vs luminosity relation of the standard model that can occur during the radiation phase of the expansion.  
Adjustment of the acceleration parameter can thus account for the leading order correction implied by an arbitrary anomalous acceleration observed at any time after the radiation phase of the Big Bang\footnotemark[5]\footnotetext[5]{This follows by the continuity of the subsequent evolution with respect to the acceleration parameter.   Higher order corrections can be derived by evolving forward, up through the $p=0$ stage of the Standard Model, the higher order corrections induced by the expanding wave perturbations at the end of the radiation phase.  The leading order corrections are then fixed by the choice of acceleration parameter, and the higher order corrections are thus a prediction of the theory.  This is a topic of the authors' current research.}.  The equations thus introduce an apparent  {\it anomalous acceleration} into the Standard Model without recourse to a cosmological constant.   

Based on this we propose to investigate whether the observed anomalous acceleration of the galaxies might be due to the fact that we are looking outward into an expansion wave.\footnotemark[6]\footnotetext[6]{The authors originally proposed the idea that a secondary expansion wave reflected backwards from the cosmic shock wave constructed in \cite{smolte1} might account for the anomalous acceleration of the galaxies, c.f. \cite{temptalk}.}  This would provide an explanation for the anomalous acceleration within classical general relativity without recourse to the {\it ad hoc} assumption of Dark Energy with its unphysical anti-gravitational properties.  Because these expanding waves have a center of expansion when $a\neq1$,  this would violate the so-called {\it Copernican Principle}, a simplifying assumption generally taken in cosmology, (c.f. \cite{cliffela}, and our discussion below in the Conclusion).  But most importantly, we emphasize that our anomalous acceleration parameter is not put in {\it ad hoc}, but rather is derived from first principles starting from a theory of non-interacting expansion waves.  The purpose of this note is to summarize our results and describe the physical interpretations. Details will appear in a forthcoming article.

\section{An Expanding Wave Coordinate System for the FRW Spacetime}
\setcounter{equation}{0}
 
 We consider the Standard Model of Cosmology during the pure radiation phase, after inflation, modeled by an FRW spacetime.   In co-moving coordinates this metric takes the form, \cite{wein},
 \begin{eqnarray}\label{FRW}
ds^2=-dt^2+R(t)^2dr^2+\bar{r}^2d\Omega^2,
\end{eqnarray}
where $\bar{r}=Rr$ measures arclength distance at fixed time $t$ and $R\equiv R(t)$ is the cosmological scale factor.  Assuming a co-moving perfect fluid with equation of state $p=\rho c^2/3$, the Einstein equations give, ( c.f. \cite{smolte}) 
 \begin{eqnarray}\label{eos}\label{hubble}
H(t)=\frac{\dot{R}}{R}=\frac{1}{2t},
 \end{eqnarray}
where $H$ is the Hubble constant\footnotemark[7].  \footnotetext[7]{Note that $H$ and $\bar{r}$ are scale independent relative to the scaling law $r\rightarrow\alpha r$, $R\rightarrow\frac{1}{\alpha} R$ of the FRW metric (\ref{FRW}), c.f.  \cite{smolte}.} 

The next theorem gives a coordinate transformation that takes (\ref{FRW}) to  the SSC form,
 
  \begin{eqnarray}\label{SSC}
ds^2=-B(\bar{t},\bar{r})d\bar{t}^2+\frac{1}{A(\bar{t},\bar{r})}d\bar{r}^2+\bar{r}^2d\Omega^2,
 \end{eqnarray}
such that $A$ and $B$ depend only on $\bar{r}/\bar{t}.$

\noindent {\bf Theorem:} Assume $p=\frac{\rho c^2}{3}$ and $k=0.$   Then the FRW metric
$$
ds^2=-dt^2+R(t)^2dr^2+\bar{r}^2d\Omega^2,
$$
under the change of coordinates

\begin{eqnarray}
\bar{t}&=&\psi_0\left\{1+\left[\frac{R(t)r}{2t}\right]^2\right\}t,\label{transt}\\
\bar{r}&=&R(t)r,\label{transr}
\end{eqnarray}
transforms to the SSC-metric
$$
ds^2=-\frac{d\bar{t}^2}{\psi_0^2\left(1-v(\xi)^2\right)}+\frac{d\bar{r}^2}{1-v(\xi)^2}+\bar{r}^2d\Omega^2,
$$
where
$$
\xi\equiv\frac{\bar{r}}{\bar{t}}=\frac{2v}{1+v^2},
$$
and $v$ is the SSC velocity given by
$$
v=\sqrt{\frac{B}{A}}\,\frac{\bar{u}^1}{\bar{u}^0}.
$$
Here $\bar{u}=(\bar{u}^0,\bar{u}^1)$ gives the $(\bar{t},\bar{r})$ components of the $4$-velocity of the sources in SSC coordinates.  (We include the constant $\psi_0$ to later account for the time re-scaling freedom in (\ref{SSC}), c.f. (2.18), page 85 of \cite{smolte}.)

We now assume $p=\rho c^2/3$ and that solutions depend only on $\xi.$ In the next section we show how the Einstein equations for metrics taking the SSC form (\ref{SSC})  reduce to a system of three ODE's.  A subsequent lengthy calculation then shows that FRW is a special solution of these equations.

\section{The Expanding Wave Equations}
\setcounter{equation}{0}

Putting the SSC metric ansatz  into MAPLE the Einstein equations $G=\kappa T$ reduce to the four partial differential equations

\begin{eqnarray}
\left\{ -r\frac{A_r}{A}+\frac{1-A}{A}\right\}&=& \frac{\kappa B}{A}r^2T^{00}
\label{one}\\
\frac{A_{t}}{A}&=&\frac{\kappa B}{A}rT^{01}\label{two}\\
\left\{ r\frac{B_r}{B}-\frac{1-A}{A}\right\}&=&\frac{\kappa }{A^2}
r^2T^{11}\label{three}\\
-\left\{ \left(\frac{1}{A}\right)_{tt}-B_r+\Phi\right\}
&=&2\frac{\kappa B}{A}r^2T^{22},\label{four}
\end{eqnarray}
where

\begin{eqnarray}
\Phi&=&\frac{B_{t}A_{t}}{2A^2B}-
\frac{1}{2A}\left(\frac{A_{t}}{A}\right)^2-\frac{B_r}{r}-\frac{BA_r}{rA}
\nonumber\\&&\ \ \ \ \ \ \ \ \ \ \ \ \ \ \ \ \ \ \ \ \ \ \ \ 
+\frac{B}{2}\left( \frac{B_r}{B}\right)^2 
-\frac{B}{2}\frac{B_r}{B}\frac{A_r}{A}.\nonumber\\
\end{eqnarray}
On smooth solutions, (\ref{one})-(\ref{four}) are equivalent to (\ref{one})-(\ref{three}) together with $Div_j T^{j1}=0$, where $Div_j T^{j1}=0$ can be written in the locally inertial form, 

\begin{eqnarray}
\left\{T_M^{01}\right\}_{,t}+\left\{ \sqrt{AB}T_{M}^{11}\right\}_{,r}
&=&-\frac{1}{2}\sqrt{AB}\left\{\frac{4}{r}T_{M}^{11}
+\frac{(1-A)}{Ar}(T_{M}^{00}-T_{M}^{11})\right.\label{div4}\\
&&\left. \ \ \ \ \ \ \ \ \ \ \ \ \ \ \ \ +\frac{2\kappa r}{A}
(T_{M}^{00}T_{M}^{11}-(T_{M}^{01})^{2})-4rT^{22}\right\},\nonumber
\end{eqnarray}
where $T_M$ denotes the Minkowski stress tensor, c.f \cite{groate}.  Assuming that $A$ and $B$ depend only on $\xi=\bar{r}/\bar{t}$, equations (\ref{one})-(\ref{three}) and (\ref{div4}) are equivalent to the three ODE's (\ref{one}), (\ref{three}), (\ref{div4}) together with the one constraint which represents the consistency condition obtained by equating $A_{\xi}$ from equations (\ref{one}) and (\ref{two}).  Setting 
$$
\kappa w\equiv\frac{\kappa}{3}\rho \bar{r}^2\,(1-v^2)^{-1},
$$
and
$$
E=\frac{\sqrt{AB}}{\xi},
$$
a (long) calculation shows that the three ODE's can be written in the form 
\begin{eqnarray}
\xi A_{\xi}&=&-E\left[\frac{4(1-A)v}{3+v^2-4vE}\right]\label{ode1}\\
\xi E_{\xi}&=&-E\left\{\left(\frac{1-A}{A}\right)\frac{2(1+v^2)-4vE}{3+v^2-4vE}-1\right\}\label{ode2}\\
\xi v_{\xi}&=&E\left[\frac{4\left(\frac{1-A}{A}\right)\left\{\cdot\right\}_N-(3+v^2-4vE)^2v^2}{2(3+v^2-4vE)\left\{\cdot\right\}_D}\right],
\label{ode3}
\end{eqnarray}
where

\begin{eqnarray}
\left\{\cdot\right\}_N&=&(v^2-1)\left\{2v^2E^2+2(v^2-3)E+(v^4-3)\right\}\label{brac1}\\
\left\{\cdot\right\}_D&=&\left\{(3v^2-1)E^2-4vE+(3-v^2)\right\},\label{brac2}
\end{eqnarray}
and the constraint becomes
$$
\kappa w=\frac{1-A}{3+v^2-4vE}.
$$
In summary we have the following theorem:
\begin{Theorem}
Assume that $A(\xi)$, $E(\xi)$ and $v(\xi)$ solve the ODE's (\ref{one})-(\ref{three}), and use the constraint (\ref{div4}) to define the density by
\begin{eqnarray}
\kappa\rho=\frac{3(1-v^2)(1-A)}{3+v^2-4vE}\,\frac{1}{\bar{r}^2}.\label{constraint1}
\end{eqnarray}
Then the metric
$$
ds^2=-B(\xi)d\bar{t}^2+\frac{1}{A(\xi)}d\bar{r}^2+\bar{r}^2d\Omega^2
$$
solves the Einstein equations with equation of state $p=\rho c^2/3.$  Moreover, the transformation (\ref{transt}), (\ref{transr}) of the FRW metric, leads to the following special SSC relations that hold on the Standard Model during the radiation phase:

\begin{eqnarray}
\xi =\frac{2v}{\psi_0(1+v^2)},\ \ 
A=1-v^2,\ \ 
E=\frac{1}{\psi_0\xi},\label{k=0values}
\end{eqnarray}
where $\psi_0$ is an arbitrary constant.   Another (long) calculation verifies directly that (\ref{k=0values}) indeed solves the equations (\ref{one})-(\ref{three}) with the constraint (\ref{constraint1}).  
\end{Theorem}
We conclude that the Standard Model of cosmology during the radiation phase corresponds to a solution of the expanding wave equations (\ref{one})-(\ref{three}) and (\ref{constraint1}) with parameter  $\psi_0$ accounting for the time-scaling freedom of the SSC metric (\ref{SSC}).  That is, the time-scaling $\bar{t}\rightarrow \psi_0\bar{t}$ preserves solutions of  (\ref{one})-(\ref{three}) and the constraint (\ref{div4})).  The next theorem states that modulo this scaling, distinct solutions of (\ref{one})-(\ref{three}), (\ref{div4}) describe a two parameter family of distinct spacetimes.

\begin{Theorem}
The replacement $\bar{t}\rightarrow\psi_0\bar{t}$ takes $A(\xi),$ $E(\xi)$ and $v(\xi)$ to $A(\xi/\psi_0),$ $E(\xi/\psi_0)$ and $v(\xi/\psi_0)$, and this scaling preserves solutions of (\ref{one})-(\ref{three}), (\ref{div4}).  Moreover, this is the only scaling law in the sense that any two solutions of (\ref{one})-(\ref{three}), (\ref{div4}) not related by the scaling $\bar{t}\rightarrow\psi_0\bar{t}$ describe distinct spacetimes.  
\end{Theorem}
Since equations (\ref{one})-(\ref{three}) admit three (initial value) parameters and one scaling law, it follows that (\ref{one})-(\ref{three}), (\ref{div4}) describes a {\em two} parameter family of distinct spacetimes. In the next section we show that by imposing regularity at the center there results a further reduction to a continuous one parameter family of expanding wave solutions,  such that one value of the parameter corresponds to the FRW metric with pure radiation sources.

\section{Leading Order Corrections to the Standard Model}
\setcounter{equation}{0} 

To obtain the leading order corrections to the FRW metric implied by equations (\ref{one})-(\ref{three}), (\ref{div4}), we linearize the equations about the FRW solution, and then expand these equations to leading order about the center $\xi=0.$  Modulo the scaling law, the resulting linearized equations admit one eigen-solution that tends to infinity as $\xi\rightarrow0$, and the other one satisfies $A(\xi)\rightarrow 1$, $B(\xi)\rightarrow 1,$ as $\xi\rightarrow0.$   Removing the singular solution leaves a two parameter family including the scaling law.   The analysis leads to the following theorem:
\begin{Theorem}
The $2$-parameter family of bounded solutions of  (\ref{one})-(\ref{three}), (\ref{div4}) that extends FRW of the Standard Model in SSC coordinates, is given in terms of the two parameters $\psi_0$ and $a$, to leading order in $\xi,$   by
\begin{eqnarray}
ds^2=-\frac{d\bar{t}^2}{\psi_0^2\left(1-\frac{a^2\psi_0^2\xi^2}{4}\right)}+\frac{d\bar{r}^2}{\left(1-\frac{a^2\psi_0^2\xi^2}{4}\right)}+\bar{r}^2d\Omega^2,\label{firstcorrection}
\end{eqnarray}
where
\begin{eqnarray}
A(\xi)&=&\left(1-\frac{a^2\psi_0^2\xi^2}{4}\right)+O(|a-1|\xi^4),\label{firstcorrectionA}\\
B(\xi)&=&\frac{1}{\psi_0^2\left(1-\frac{a^2\psi_0^2\xi^2}{4}\right)}+O(|a-1|\xi^4),\label{firstcorrectionB}
\end{eqnarray}
and
\begin{eqnarray}
v(\xi)=\frac{\psi_0\xi}{2}+O(|a-1|\xi^3).\label{firstcorrectionv}
\end{eqnarray}
Here $\psi_0$ is the time-scaling parameter, $a=1$ corresponds to FRW, and $a\neq1$ introduces a new acceleration parameter which gives the leading order perturbation of FRW.  (Note that (\ref{firstcorrectionv}) implies that the velocity $v$ is independent of $a$ to leading order in $\xi$.)  
\end{Theorem}
In light of (\ref{k=0values}), when $a=1$, (\ref{firstcorrection}) reduces exactly to the  FRW metric
\begin{eqnarray}
A(\xi)&=&\left(1-\frac{\psi_0^2\xi^2}{4}\right),\\\label{firstcorrectionA}
B(\xi)&=&\frac{1}{\psi_0^2\left(1-\frac{\psi_0^2\xi^2}{4}\right)}.\label{firstcorrectionB}
\end{eqnarray} 
Now the SSC coordinate representation of FRW depends only on $H$ and $\bar{r}$, both of which are invariant under the scaling $r\rightarrow\alpha r$, $R\rightarrow R/\alpha$  of the FRW metric (\ref{FRW}).  It follows that the SSC representation of FRW is independent of $\alpha,$ and therefore independent of our choice of scale for $R(t)$.  Thus without loss of generality, we can take $\psi_0=1$, and we can also assume throughout that the FRW metric is scaled exactly so that 
\begin{eqnarray}
R(t)=\sqrt{t},\label{Roft}
\end{eqnarray}
c.f. (\ref{FRW}) and  \cite{smolte}.  We conclude that to leading order, the $1$-parameter family of expanding wave perturbations of the FRW metric is given by
\begin{eqnarray}
ds^2=-\frac{d\bar{t}^2}{\left(1-\frac{a^2\xi^2}{4}\right)}+\frac{d\bar{r}^2}{\left(1-\frac{a^2\xi^2}{4}\right)}+\bar{r}^2d\Omega^2,\label{metrica}
\end{eqnarray}
with errors of order $|a-1|\xi^4$, ($a=1$ corresponding {\em exactly} to FRW),  and the velocity is given to leading order  by
\begin{eqnarray}
v=\frac{\xi}{2},\label{v}
\end{eqnarray}
{\em independent of $a$}, with errors of order $|a-1|\xi^3$.

\section{Comoving Coordinates and Comparison with the Standard Model}
\setcounter{equation}{0} 

Since (\ref{v}) is independent of $a,$ it follows that even when $a\neq1,$ the inverse of the transformation (\ref{transt}), (\ref{transr}) gives, to leading order in $\xi$, a co-moving coordinate system for (\ref{metrica}) in which we can compare the Hubble constant and redshift vs luminosity relations for (\ref{metrica}) when $a\neq 1$ to the Hubble constant and redshift vs luminosity relations for FRW as measured by (\ref{FRW}) and (\ref{hubble}).
\begin{Theorem}
Take $\psi_0=1,$ and set
$
\zeta=\bar{r}/t.
$
Then the inverse of the coordinate transformation (\ref{transt}), (\ref{transr}) maps (\ref{metrica}) over to $(t,r)$-coordinates as
\begin{eqnarray}
ds^2=F_a(\zeta)^2\left\{-dt^2+tdr^2\right\}+\bar{r}^2d\Omega^2,\label{metricb}
\end{eqnarray}
where
\begin{eqnarray}
F_a(\zeta)^2=1+(a^2-1)\frac{\zeta^2}{4}+O(|a-1|\zeta^4),\label{errorF}
\end{eqnarray}
and the SSC velocity $v$ in (\ref{v}) maps to the $(t,r)$-velocity
\begin{eqnarray}
\tilde{v}=O(|a-1|\zeta^3).\label{vb}
\end{eqnarray}
The errors in (\ref{errorF}) and (\ref{vb}) are written in terms of the co-moving coordinate variable $\zeta$, which by (\ref{transt}), (\ref{transr}) satisfies 
$\zeta=O(\xi)\ as\ \xi\rightarrow0,
$
and $\bar{r}=R(t)r$, the area of the spheres of symmetry, exactly measures arclength at $t=const.$ in FRW when $a=1.$ 
\end{Theorem}
Note that $\zeta=\bar{r}/t$ is a natural dimensionless perturbation parameter that has a physical interpretation in $(t,r)$-coordinates because, (assuming $c=1$ or $t\equiv ct$), $\zeta$ ranges from $0$ to $1$ as $\bar{r}$ ranges from zero to the horizon distance in FRW, (approximately the Hubble distance $c/H$), a measure of the furthest one can see from the center at time $t$ units after the Big Bang, \cite{wein};  that is,
\begin{eqnarray}\zeta&\approx&\frac{Dist}{Hubble\ Length}.\label{horizon}
\end{eqnarray}
Thus expanding in $\zeta$ gives an expansion in the fractional distance to the Hubble length, c.f. \cite{smolte}.
Note also that when $a=1$ we obtain the FRW metric (\ref{FRW}), where we have used $R(t)=\sqrt{t},$ c.f. (\ref{Roft}).  

Now for a first comparison of the relative expansion at $a\neq1$ to the expansion of FRW, define {\it the Hubble constant at parameter value a}, by
$$H_a(t,\zeta)=\frac{1}{R_a}\frac{\partial}{\partial t}R_a,$$
where
$$
R_a(t,\zeta)=F_a(\zeta)\sqrt{t},
$$
equals the square root of the coefficient of $dr^2$ in (\ref{metricb}).
Then one can show

$$H_a(t,\zeta)=
\frac{1}{2t}\left\{1-\frac{3}{8}(a^2-1)\zeta^2+O\left(|a^2-1|\zeta^4\right)\right\}.$$
We conclude that the fractional change in the Hubble constant due to the perturbation induced by expanding waves $a\neq1$ relative to the FRW of the Standard Model $a=1$, is given by
$$\frac{H_a-H}{H}=\frac{3}{8}(1-a^2)\zeta^2+O\left(|a^2-1|\zeta^4\right).$$

A different mapping to $(t,r)$ coordinates gives additional insight into the geometry of the spacetime metric (\ref{metricb}) when $a\neq1.$
Namely, if we extend the FRW $(t,r)$ coordinates to $a\neq1$ by
\begin{eqnarray}
\bar{t}&=&\left\{1+\frac{a^2\zeta^2}{4}\right\}t,\label{transthat}\\
\bar{r}&=&t^{a/2}r,\label{transrhat}
\end{eqnarray}
then one can show that in this case the metric (\ref{metrica}) transforms to
\begin{eqnarray}
ds^2=-dt^2+t^adr^2+\bar{r}^2d\Omega^2+a(1-a)\zeta dtd\bar{r}.\label{metricc}
\end{eqnarray}
The coordinate system $(t,r)$ as defined by (\ref{transthat}), (\ref{transrhat}) does not define a comoving coordinate system  for (\ref{metrica}) to leading order, as does (\ref{transt}), (\ref{transr}).  However,  (\ref{metricc}) takes the form of a $k=0$ Friedmann-Robertson-Walker metric with a small correction to the scale factor, ($R_a(t)=t^{a/2}$ instead of $R(t)=t^{1/2}$), and a small corrective mixed term.  In particular, the time slices $t=const.$ in (\ref{metricc}) are all flat space ${\mathcal R}^3,$ as in FRW, and the $\bar{r}=const$ slices agree with the FRW metric modified by scale factor $R_a(t).$  Thus (\ref{metricc}) exhibits many of the flat space properties characteristic of FRW.

\section{Redshift vs Luminosity Relations}
\setcounter{equation}{0} 

In this section we obtain the first order corrections to the redshift vs luminosity relations of FRW, as measured by an observer positioned at the center $\zeta=0$ of the expanding wave spacetimes described by the metric  (\ref{metrica}) when $a\neq1.$   (Recall that $\zeta\equiv\bar{r}/t$ measures the fractional distance to the horizon, c.f. (\ref{horizon}).)  The physically correct coordinate system in which to do the comparison with FRW ($a=1$) should be comoving with respect to the sources.  Thus we restrict to the coordinates $(t,r)$ defined by (\ref{transt}), (\ref{transr}), in which our one parameter family of expanding wave spacetimes are described, to leading order in $\zeta$, by the metric (\ref{metricb}).  Note that (\ref{metricb}) reduces {\em exactly} to the FRW metric when $a=1,$ c.f. (\ref{FRW}),  (\ref{Roft}).  For our derivation of the redshift vs luminosity relation for (\ref{metricb}) we follow the development in \cite{gronhe}.

To start, assuming radiation is emitted by a source at time $t_e$ at wave length $\lambda_e$ and received at $\zeta=0$ at later time  $t_0$ at wave length $\lambda_0,$  define
\begin{eqnarray}
L&\equiv&Absolute\ Luminosity=\frac{\rm Energy\ Emitted\ by\ Source}{\rm Time},\label{lum1}\\
d_{\ell}&\equiv& Luminosity\ Distance=\left(\frac{L}{4\pi\ell }\right)^{1/2},\label{lum2}\\
\ell&\equiv& Apparent\ Luminosity=\frac{\rm Power\ Recieved}{\rm Area},\label{lum3}\\
z&\equiv& Redshift\ Factor=\frac{\lambda_0}{\lambda_e}-1.\label{lum4}
\end{eqnarray}
Then using two serendipitous properties of the metric (\ref{metricb}), namely, the metric is diagonal in co-moving coordinates, and there is no $a$-dependence on the sphere's of symmetry, it follows that the arguments in \cite{gronhe}, Section 11.8, 
 can be modified to give the following result:

\begin{Theorem}
The redshift vs luminosity relation, as measured by an observer positioned at the center $\zeta=0$ of the spacetime described by metric (\ref{metrica}), is given to leading order in redshift factor $z$ by
\begin{eqnarray}
d_{\ell}=2\sqrt{t_0}z\left\{1+\frac{3}{2}\left(a^2-1\right)\frac{z}{1+z}\right\}+O\left(|a-1|z^3\right),\label{redvslum}\label{redvslumFRW}
\end{eqnarray}
where we used the fact that $z$ and $\zeta$ are of the same order as $\zeta\rightarrow0.$
\end{Theorem}
Note that when $a=1$, (\ref{redvslum}) reduces to the well known FRW linear relation, 
\begin{eqnarray}
d_{\ell}=2\sqrt{t_0}z,\nonumber
\end{eqnarray}
correct for the radiation phase of the Standard Model, \cite{gronhe}.
Thus the bracket in (\ref{redvslum}) gives the leading order quadratic correction to the redshift vs luminosity relation implied by the change in the Hubble expansion law corresponding to expanding wave perturbations of the FRW spacetime when $a\neq1.$  Since $(a^2-1)$ appears in front of the leading order correction in (\ref{redvslum}), it follows that the leading order part of any anomalous correction to the redshift vs luminosity relation of the Standard Model, observed at a time after the radiation phase, can be accounted for by suitable adjustment of parameter $a$.   In particular, note that the leading order corrections in (\ref{redvslum})  imply a blue-shifting of radiation relative to the Standard Model, as observed in the supernova data, when $a>1,$ \cite{gronhe}.

\section{Concluding Remarks}
\setcounter{equation}{0} 

We have constructed a one parameter family of general relativistic expansion waves which at a single parameter value, reduces to the FRW spacetime, the Standard Model of Cosmology during the radiation epoch.  The discovery of this family is made possible by a remarkable coordinate transformation that maps the FRW metric in standard comoving coordinates, over to Standard Schwarzschild coordinates (SSC) in such a way that all quantities depend only on the  single self-similar variable $\xi=\bar{r}/\bar{t}$.  Note that it is not evident from the FRW metric in standard comoving coordinates that self-similar variables even exist, and if they do exist, by what ansatz one should extend the metric in those variables to obtain nearby self-similar solutions that solve the Einstein equations exactly.   The main point is that our coordinate mapping to SSC form, explicitly identifies the self-similar variables as well as the metric ansatz that together accomplish such an extension of the metric.

The self-similarity of the FRW metric in SSC suggested the existence of a reduction of the SSC Einstein equations to a new set of ODE's in  $\xi.$  Deriving this system from first principles then establishes that the FRW spacetime does indeed extend to a three parameter family of expanding wave solutions of the Einstein equations.  This three parameter family reduces to an (implicitly defined) one parameter family by removing a scaling invariance and imposing regularity at the center.  The remaining parameter $a$ changes the expansion rate of the spacetimes in the family, and thus we call it the {\it acceleration parameter}.   Transforming back to comoving coordinates, the resulting one parameter family of metrics is amenable to the calculation of a redshift vs luminosity relation, to second order in the redshift factor $z,$ leading to the relation (\ref{redvslum}).  It follows by continuity that the leading order part of any anomalous correction to the redshift vs luminosity relation of the Standard Model observed {\em after} the radiation phase, can be accounted for by suitable adjustment of parameter $a$.  

These results suggest an interpretation that we might call a {\it Conservation Law Scenario} of the Big Bang.  That is, it is well known that highly oscillatory interactive solutions of conservation laws decay in time to non-interacting waves, (shock waves and expanding waves), by the mechanism of shock wave dissipation.  The subtle point is that even though dissipation terms are neglected in the formulation of the equations, there is a canonical dissipation and consequent loss of information due to shock wave interactions that drive solutions to non-interacting wave patterns.  (This viewpoint is well expressed in the celebrated works \cite{lax,glim,glimla}).  Since the one fact most certain about the Standard Model is that our universe arose from an earlier hot dense epoch in which all sources of energy were in the form of radiation, one might reasonably conjecture that decay to a non-interacting expanding wave occurred during the radiation phase of the Standard Model, via the highly nonlinear evolution driven by the large sound speed present when $p=\rho c^2/3$.   Our analysis has shown that FRW is just one point in a family of non-interacting expanding waves, and as a result we conclude that some further explanation is required as to why, on some length scale, decay during the radiation phase of the Standard Model would not proceed to a member of the family satisfying $a\neq1.$  If decay to $a\neq1$ did occur, then the galaxies that formed from matter at the end of the radiation phase, (some $379,000$ years after the Big Bang), would be displaced from their anticipated positions in the Standard Model at present time, and this displacement would lead to a modification of the observed redshift vs luminosity relation.  In principle such a mechanism could account for the anomalous acceleration of the galaxies as observed in the supernova data.  Of course, if $a\neq1$, then the spacetime has a center, and this would violate the so-called {\it Copernican Principle}, a simplifying assumption generally accepted in cosmology, (c.f. the discussions in \cite{temptalk} and \cite{copihudjgd}).  As a consequence, if the earth did not lie within some threshold of the center of expansion, the expanding wave theory would imply large angular variations in the observed expansion rate.\footnotemark[8]\footnotetext[8]{The size of the center, consistent with the angular dependence that has been observed in the actual supernova and microwave data, has been estimated to be about $15$ megaparsecs, approximately the distance between clusters of galaxies, roughly $1/200$ the distance across the visible universe, c.f. \cite{copihudjgd,wein}.} In any case, the expanding wave theory presented here can in principle be tested.  For this, one must evolve the relation (\ref{redvslum}), valid at the end of the radiation phase, up through the $p=0$ stage to present time in the Standard Model,  thereby obtaining the exact value of $a$ that gives the leading order correction to redshift vs luminosity relation observed in the supernova data.  Then a derivation of the next order correction to (\ref{redvslum}) at that $a$-value, evolved up through the $p=0$ stage to present time in the Standard Model,  would make a {\em prediction} that could be compared with an accurate plot of the supernove observations of redshift vs luminosity.   (This is a topic of the authors' current research.)

To summarize, the expanding wave theory could in principle give an explanation for the observed anomalous acceleration of the galaxies within classical general relativity, with classical sources.  In the expanding wave theory, the so-called anomalous acceleration is not an acceleration at all, but is a correction to the Standard Model due to the fact that we are looking outward into an expansion wave.  The one parameter family of non-interacting general relativistic expansion waves derived here, are all equally possible end-states that could result after dissipation by wave interaction during the radiation phase of the Standard Model is done; and when $a\neq1$ they introduce an anomalous acceleration into the Standard Model of cosmology.   Unlike the theory of Dark Energy, this provides a possible explanation for the anomalous acceleration of the galaxies that is not {\it ad hoc} in the sense that it is derivable exactly from physical principles and a mathematically rigorous theory of expansion waves.  In particular, this explanation does not require the {\it ad hoc} assumption of a universe filled with an as yet unobserved form of energy with anti-gravitational properties\footnotemark[9]\footnotetext[9]{I.e., the standard physical interpretation of the cosmological constant.} in order to fit the data.  The idea that the anomalous acceleration might be accounted for by a local under-density in a neighborhood of our galaxy was expounded in the recent paper \cite{cliffela}.   Our results here might then give an accounting of the source of such an under-density.  

In conclusion, these new expanding wave solutions of the Einstein equations provide a new paradigm to test against the Standard Model.  Moreover, even if these new general relativistic expansion waves do not in the end explain the anomalous acceleration of the galaxies, their presence represents an instability in the Standard Model in the sense that an explanation is required as to why oscillations have to settle down to $a=1$ expansions instead of $a\neq1$ expansions, (either locally or globally), during the radiation phase of the Big Bang.

\end{document}